\documentclass[slac_one]{revtex4}
\usepackage{graphicx}
\usepackage{fancyhdr}
\begin{document}

\title{Heavy Quark Production at the Tevatron} 

\author{Sally Seidel}
\affiliation{Department of Physics and Astronomy, University of New Mexico,
Albuquerque, NM 87131 \\
\ \\
for the CDF Collaboration}

\begin{abstract}
Results are presented from four CDF analyses involving heavy quark production
in proton-antiproton collisions at center of mass energy 1.96 TeV.  The shapes
of b-jets are found to be broader than inclusive predictions and broader than
both PYTHIA and HERWIG defaults.  A measurement of the production cross
section for $\psi(2S)$ is consistent with Run 1 results and with theoretical
predictions associated with parton distribution function energy dependence.
The inclusive b-jet production cross section is also consistent with
theoretical predictions over six orders of magnitude.  The $b{\overline b}$
differential production cross section is compared to several theoretical models
and found to be best described by MC@NLO + JIMMY.
\end{abstract}

\maketitle

\thispagestyle{fancy}

\section{Introduction}
We report on four measurements by the CDF Experiment~\cite{cdf} of processes 
involving
heavy quark production.  All use data recorded at center of mass energy
1.96 TeV at the Fermilab Tevatron Collider during Run II. 

\section{Measurement of $b$-jet Shapes in Inclusive Jet Production}

The structure of jets derives from the gluon emissions from the primary
parton involved.  In heavy quark jets, the quark decay must be modeled as 
well as the underlying event.  Multi-gluon emission is difficult
to calculate so is often approximated by parton shower models.  Jet shape
is known to depend upon whether the primary parton is a quark or a gluon, and
it is also expected to depend upon flavor.  It is expected to depend as well
upon the production mechanism; for example, the $b$ and $\overline b$ from
gluon splitting are expected to be often in the same jet~\cite{frixione}, 
producing a broader
jet than does flavor creation.  Measurement of jet shapes casts light on all
these aspects of jet evolution and heavy flavor production.

To address the question of whether the fraction of $b$-jets originating from
gluon splitting, and its evolution, is well described by contemporary models,
we define the jet shape, a measure of the fraction of total jet transverse
momentum $p_{\rm T}$, inside a given radius in the space of rapidity $y$ and
azimuthal angle $\phi$.  This is
$\Psi(r/R)\equiv \langle {p_T(0\rightarrow r)\over p_T(0 \rightarrow R)}\rangle$,
the fraction of total $p_{\rm T}$ in cone $R$ carried by particles in subcone
$r$, averaged over an ensemble of jets, normalized to 1, and defined such
that $\Psi(0)=0$ and particles outside the cone are excluded.

The $b$-jet event selection begins with triggers based solely on calorimeter transverse energy $E_{\rm T}$.  Trigger
Level 1 requires one calorimetric trigger tower above a threshold between 5
and 10 GeV.  Level 2 seeks clusters about the Level 1 towers among adjacent
towers above 1 GeV and requires that at least one cluster exceed a threshold
between 15 and 90 GeV.  Level 3 applies the Run 1 cone algorithm~\cite{jetclu}
and imposes $E_{\rm T}^{\rm min}$ greater than 20 to 100 GeV.  The jet
minimum $p_{\rm T}$ requirement is set to ensure that only events for which
the trigger is $\ge 99\%$ efficient are used.  Offline the jets are
reconstructed by the Midpoint Cone Algorithm~\cite{midpoint}.  The selection
of b-jets is then enhanced by requiring a secondary vertex.  Because $b$'s
tend to be found near the jet axis, the search for them uses a cone of
radius $R=0.4$ in $(y,\phi)$ space.  Tracks are ranked by reconstruction 
quality, including the contribution from the distance $d_0$ of closest 
approach to the primary vertex.  The algorithm next attempts
to reconstruct a secondary
vertex, beginning with the highest quality track.  If this succeeds, a cut
is placed on the significance of the two-dimensional projection along the
jet axis of the distance between the primary and secondary vertices.
General event quality cuts follow, including a rejection of multiple 
interactions (there must be one and only one primary vertex with $|z|<50$ cm),
a rejection of cosmics based on the significance of the missing $E_T$,
and a requirement of jet centrality: $|y_{\rm jet}| \le 0.7$.  The
jet $p_T$ is corrected to the hadron level by matching hadron and
calorimeter level jets in Monte Carlo, a correction that increases the
$p_T$ by 20\% to 10\% with no change in jet shape.

The requirement of the secondary vertex tag biases the measured jet shapes
by demanding clean, well-defined tracks.  An $r$-dependent correction for
this is needed.  There are additionally corrections for the presence of
non-$b$ jets misidentified as $b$-jets, corrections for detector effects (for
example, the number of calorimeter towers scales with $p_{\rm T}$), and a
correction for the presence of a second $b$ quark in the jet.  The data
are fitted to two-$b$ and one-$b$ templates separately.  The final function
to be fitted is given by $\Psi_{\rm had}^{b}=C^{\rm had}(r/R) \cdot
{\Psi_{\rm det}^{\rm tag}(r/R)-(1-p_b)b_{\rm non-b}(r/R)\Psi_{\rm det}^{\rm non-b}(r/R) \over p_bb_b(r/R)}$,
where $C^{\rm had}$ is the unfolding factor $\Psi_{\rm had}^b(r/R)/\Psi_{\rm det}^b(r/R)$,
$\Psi_{\rm det}^{\rm tag}$ is the measured jet shape for the tagged sample,
$\Psi_{\rm det}^{\rm non-b}$ is the measured inclusive jet shape, $p_b$ is
the purity of $b$'s in the tagged jet sample, and $b_b$ and $b_{\rm non-b}$
correct for biases to the jet shape arising from the secondary vertex 
requirement.
The data are compared to predictions by PYTHIA, with Tune A~\cite{pythia},
and HERWIG v 6.506~\cite{herwig}.  The principle systematics 
are associated
with the choice of Monte Carlo model, the effect of the calorimeter model, the
jet energy calibration, and assumptions about the $c$-quark content of the
jets.

\begin{figure}
\includegraphics[width=135mm]{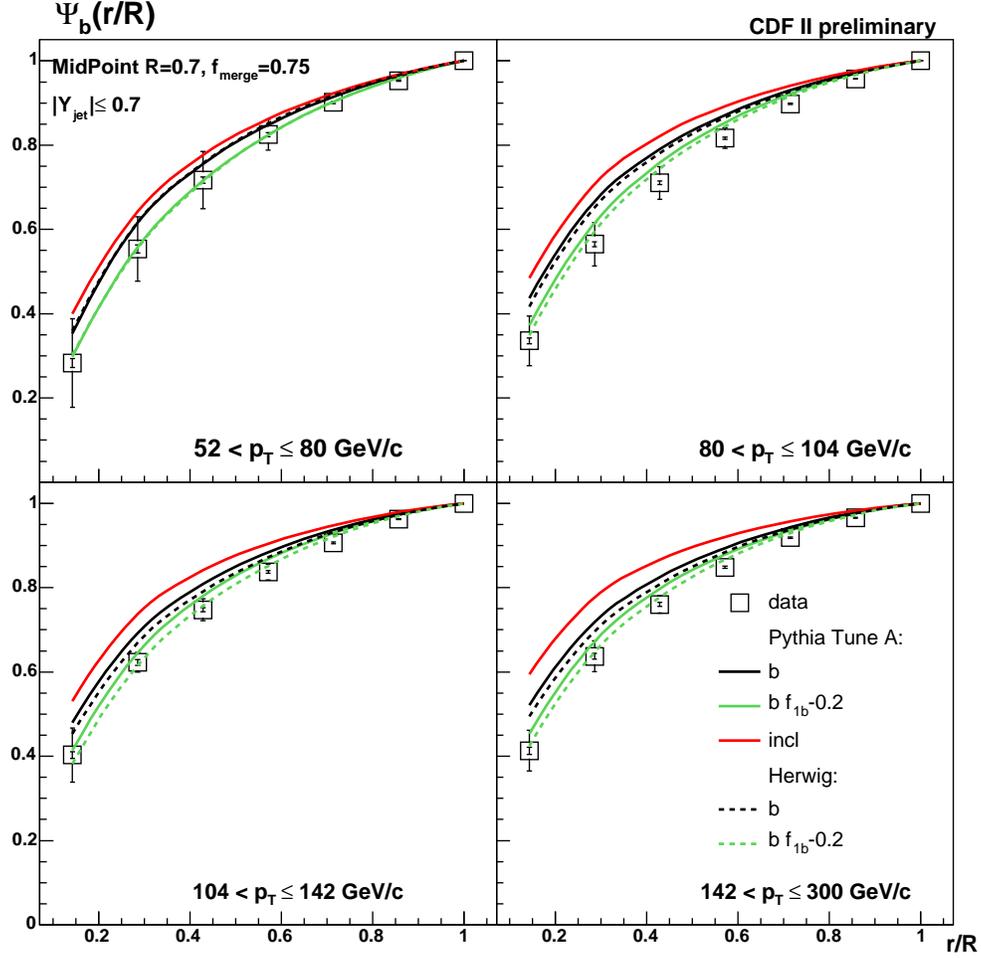}
\caption{The measured integrated $b$-jet shapes for four $p_{\rm T}$ bins.
The data are shown as open squares where the error bars represent the 
statistical and total uncertainties.  The statistical uncertainties are smaller
than the squares.  The data are compared to PYTHIA Tune A (solid lines) and 
HERWIG (dashed lines).  The uppermost solid curve is the inclusive prediction;
the next solid curve, and the uppermost dashed curve, are 
the predictions with the default $f_{1b}$ fraction; and the lowest
solid curve and lowest dashed curve are the predictions
with the default 1$b$ fraction diminished by 20\%.\label{unfold2}}
\end{figure}

The results of the study~\cite{result}, for integrated luminosity 
300 pb$^{-1}$, are shown
in Figure~\ref{unfold2}, for 4 bins in $p_{\rm T}$.  In each bin, the data
are compared to the PYTHIA prediction for inclusive jets as well as the
PYTHIA and HERWIG predictions for single $b$ production and for single $b$
production with the default fraction diminished by 20\%.  One sees that the
data are incompatible with the inclusive prediction: the jet shapes are
evidently influenced by the presence of heavy quarks, and $\sl b$-jets are broader.
One also sees that the data jets are broader than the PYTHIA and HERWIG 
defaults; the leading order models underestimate the fraction of $b$'s from
gluon splitting and the fraction of jets with two $b$'s.

\section{Measurement of the Production Cross Section of 
the ${\mathbf \psi(2S)}$}

The mechanism for producing heavy vector mesons in hadron collisions is not
well understood.  CDF Run I data on production cross sections for prompt 
$J/\psi$ and $\psi(2S)$ were one to two orders of magnitude above predictions
by color singlet models.  Subsequent theoretical efforts~\cite{nrqcd} with
adjusted production matrix elements match the cross section but predict 
increasing transverse polarization with production $p_{\rm T}$ that is not
confirmed by the data~\cite{abulencia}.  New approaches~\cite{kt, tower} have
been proposed.  The $\psi(2S)$ is a good testing ground for studying 
charmonium hadroproduction as there are no significant charmonium states above
it to produce feeddown.  The goal of this study is a measurement of
$\sigma(p{\overline p}\rightarrow \psi(2S))\cdot BR(\psi(2S)\rightarrow\mu^+\mu^-)$
for $2\le p_{\rm T} \le 30$ GeV/c.

The data selection begins with muons reconstructed in four layers of the 
central tracking chamber and matched to three to four hit tracks
in the muon detector.  Events that pass the dimuon trigger, which
requires two opposite sign tracks, each with significant $p_{\rm T}$, are
required to have three hits in the silicon vertex detector SVX II.
The $\psi(2S)$ mass and lifetime are fit with a lifetime function,
and requirements are placed on the dimuon mass, 
$3.5 < m(\mu\mu)<3.8$ GeV/c$^2$,
and rapidity $|y(\mu\mu)|<0.6$.

CDF separates the signal from background, and the prompt from the $b$-decay 
$\psi$'s, with an unbinned maximum likelihood fit in candidate mass and proper
decay length $ct$.  The mass separates the signal from the background.  The
signal is modeled with the Crystal Ball Function~\cite{cbf}, a Gaussian core
with a low side tail.  The mass background is modeled with a first order
polynomial.  The prompt signal is separated from the feeddown with a $ct$-fit.
The prompt signal is given by a double Gaussian centered on zero.  The
long-lived signal is modeled by an exponential convoluted with a Gaussian.
The lifetime background is modeled by a prompt double Gaussian plus symmetric,
positive-$ct$, and negative-$ct$ long-lived components.  The likelihood 
function is given by $L=f_sP_s^{\rm mass}(f_pP_p^{ct}+(1-f_p)P_{E \otimes G}^{ct})
+(1-f_s)P_{\rm bkg}^{\rm mass}(f_{\rm sym}P_{\rm sym}^{ct}+f_+P_+^{ct}+
f_-P_-^{ct}+(1-f_{\rm sym}-f_+-f_-)P_p^{ct})$, where the $f$'s are fractions
of signal ($s$, from the total number of candidates in the fit), prompt ($p$),
symmetric long-lived background (sym), positive-$ct$ long-lived background (+),
and negative-$ct$ long-lived background(-).  The $P$'s are probability density
functions for signal mass $(P_s^{\rm mass})$, linear mass background 
$(P_{\rm bkg}^{\rm mass})$, prompt signal double Gaussian proper time
$(P_p^{ct})$, an exponential convolved with a Gaussian for $b$-decay signal 
$(P^{ct}_{E \otimes G})$, and lifetime backgrounds
($P_{\rm sym}^{ct}$, $P_+^{ct}$, and $P_-^{ct}$).  

The efficiency calculation is given by the product of six
terms whose values range from 95.3\% to 99.9\%. The acceptance calculation
is complicated by the fact that acceptance depends upon polarization per
$p_{\rm T}$ bin, yet CDF polarization data on the $\psi(2S)$ are too weak
for direct application.  The $J/\psi$ polarization, for which strong CDF
data exist, are, however, expected to correlate highly with prompt $\psi(2S)$
polarization, so after confirming that the $J/\psi$ and $\psi(2S)$ polarization
data are consistent, CDF applies the $J/\psi$ polarization values to the 
$\psi(2S)$ $p_{\rm T}$ bins.  Monte Carlo $\psi(2S)$ samples are generated with
fixed polarizations at the extreme values of 0 and -1 and with flat
distributions in $p_{\rm T}$, $\eta$, and $\phi$.  These samples are simulated
in the CDF detector and reconstructed.  We compute the product of
geometrical acceptance and trigger efficiency,
$A={N^{\rm rec}(p_{\rm T})\times(N^{\rm eff}(p_{\rm T})/N^{\rm rec}(p_{\rm T}))\over N^{\rm gen}(p_{\rm T})}$,
where for $N^{\rm gen}$ generated events, $N^{\rm rec}$ survive geometric and
reconstruction requirements and $N^{\rm eff}$ survive the trigger.  $A$ ranges
from 0.0053 to 0.2548.  We interpolate the data to intermediate polarization
values predicted from the $J/\psi$ data.  The principle systematic uncertainties
on this measurement are associated with the luminosity, the reconstruction
efficiency, the dimuon trigger efficiency, and the $\psi(2S)$ polarization.
The variation of the prescale during the course of collecting
integrated luminosity 1.1 fb$^{-1}$ produces an effective integrated luminosity
of 954 pb$^{-1}$.  

\begin{figure}
\includegraphics[width=135mm]{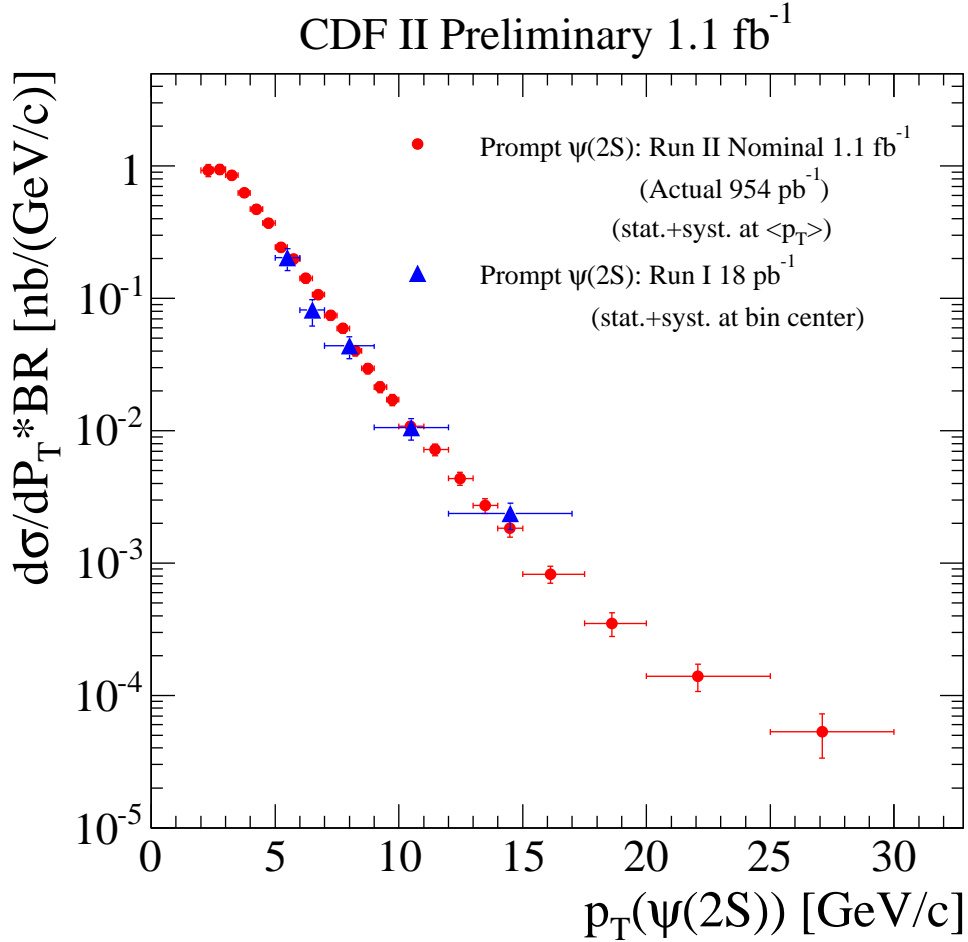}
\caption{The prompt production cross section versus $p_{\rm T}$ for data from
CDF Run I (triangles) and Run II (circles).\label{gdiff3}}
\end{figure}

The resulting 
prompt $\psi(2S)$ differential
cross section
is shown in Figure \ref{gdiff3}.
One
sees that in all three cases, the Run I result is confirmed with an order
of magnitude more statistics.  The measured value of 
$\sigma(p{\overline p}\rightarrow \psi(2S))\cdot BR(\psi(2S)\rightarrow\mu^+\mu^-)$ is $0.68\pm 0.01 \pm 0.06$~nb.  The ratio of the
Run II to 
the Run I result, for identical ranges in $p_{\rm T}$, is $18 \pm 19\%$.  The
theoretical prediction~\cite{anikeev} for this ratio, including the energy
dependence of the parton distribution functions, is $14 \pm 8\%$.

\section{Study of Inclusive $b$-jet Production}

Measurements of $b$ quark production in hadronic collisions test perturbative
QCD.  The CDF Run I inclusive $B$ meson cross section measurement~\cite{b}
motivated theoretical developments beyond next-to-leading order~\cite{nlo}.
Continued comparisons, at higher $p_{\rm T}$ and with increased statistics, may
motivate further improvements.

\begin{figure}
\includegraphics[width=135mm]{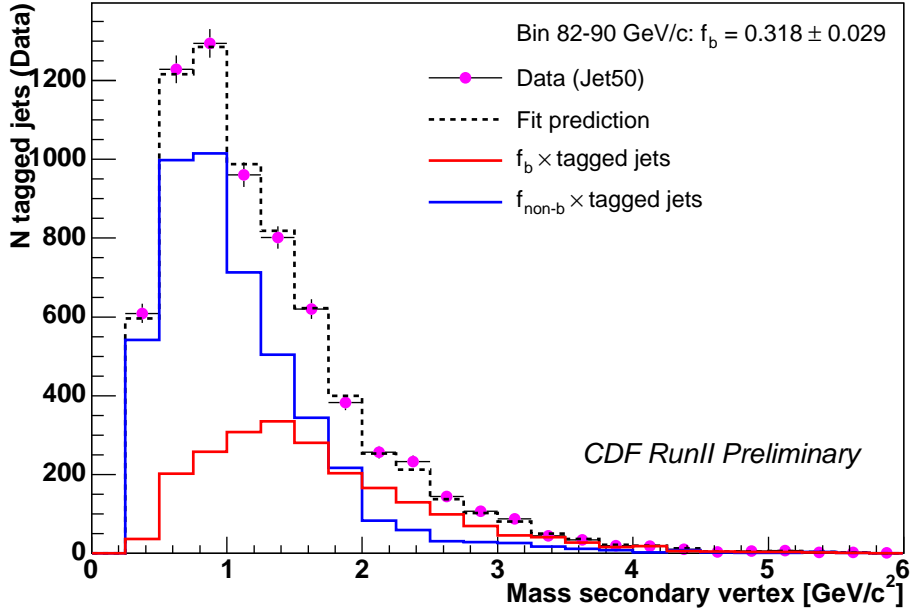}
\caption{The measured secondary vertex mass distribution for jets with 
corrected jet $p_{\rm T}$ in the range 80 to 92 GeV/c, compared to the result
from the fit to Monte Carlo templates for $b$ and non-$b$ jets. \label{bmass1}}
\end{figure}

This study considers events whose primary vertex has $|z|<50$ cm, whose
missing $E_{\rm T}$ is not significant, and whose kinematical variables lie
in the range $38<p_{\rm T}^{\rm jet}<400$ GeV/c and $|y^{\rm jet}|<0.7$,
with jets reconstructed with the Midpoint Cone Algorithm and heavy
flavor jets tagged via their secondary vertices.  The jet energy corrections
are based on minimum bias events which measure the $p_{\rm T}$ deposited
in the calorimeter, as a function of the number of primary vertices, and
indicate a value of $0.93 \pm 0.14$ GeV/c per extra vertex.  Monte Carlo
events from PYTHIA 6.203, with Tune A and the CTEQ5L parton distribution
function
set, are used to correct the measured energy for effects of partially
instrumented regions and calorimeter nonlinearities.  Typical unfolding
factors range from 1.6 to 2.1.  The $b$ quark content of the jet is inferred
from the shape of the invariant mass of all charged tracks attached to the
secondary vertex, as shown in Figure~\ref{bmass1}.  The measurement is 
compared to the theoretical prediction described in \cite{frixione2}.

\begin{figure}
\includegraphics[width=135mm]{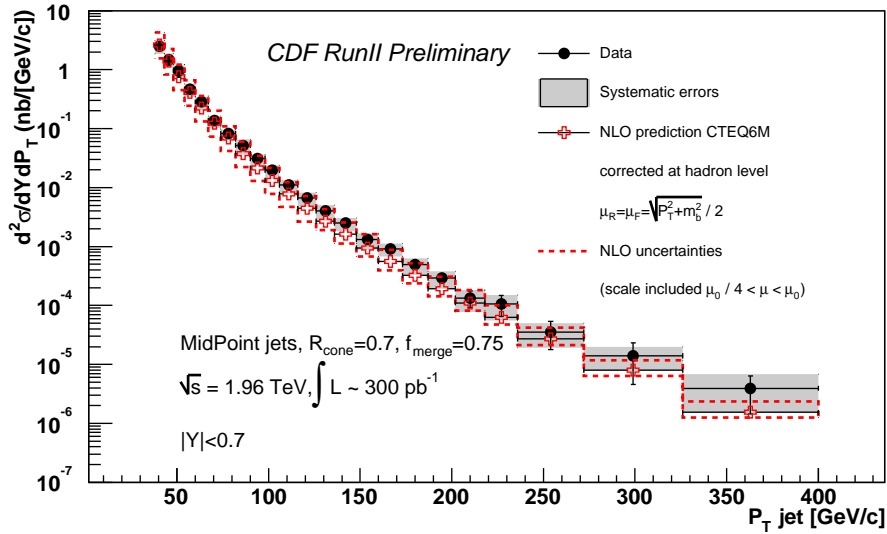}
\caption{The measured inclusive $b$-jet cross section (filled circles) as a 
function of $p_{\rm}^{\rm jet}$ compared with the next-to-leading order 
perturbative QCD prediction (empty crosses).  The shaded band represents
the systematic uncertainty on the data and the dashed band represents the
uncertainty on the theory. \label{bcross1}}
\end{figure}

The measured differential cross section, 
${d^2\sigma_{b-jet}\over{dp_{\rm T}dy}}=
{N_{\rm tagged}f_b \over \epsilon_{\rm b-tag}\Delta y^{\rm jet} \Delta p_{\rm T}^{\rm jet} \int L dt}$, is shown in Figure~\ref{bcross1} for integrated luminosity
300 pb$^{-1}$.  Here $N_{\rm tagged}$ is the number of tagged jets in each
$p_{\rm T}$ bin, $\Delta p_{\rm T}$ is the bin size, $f_b$ is the fraction
of jets in the tagged sample (this ranges from 0.35 at low $p_{\rm T}$ to
0.14 at high), $\epsilon_{\rm b-tag}$ is the $b$-tagging efficiency, 
$\Delta y^{\rm jet}$ is the jet rapidity range, and $L$ is the luminosity.
The measured cross section is consistent with the theoretical prediction 
throughout the full six orders of magnitude of range in $p_{\rm T}$.  It 
exhibits a strong dependence on scale $\mu$, however, which suggests that
higher orders may make large contributions.  The largest
systematic uncertainties derive from the luminosity, the jet energy scale,
the jet energy resolution, the unfolding process, the $b$-tagging efficiency,
and the fraction of $b$-jets, and produce a total systematic uncertainty
that ranges from $\pm 25\%$ at low $p_{\rm T}$ to $\pm 70\%$ at high
$p_{\rm T}$.

\section{Study of the Production of $b{\overline b}$ Dijets}

Momentum conservation requires that the azimuthal angle $\phi$ between a $b$
and $\overline b$ produced at lowest order must be $180^\circ$.  Higher order
QCD processes produce additional partons in the final state, modifying the
range of allowed azimuthal angle difference, $\Delta \phi$.  Order 
$\alpha_s^3$ diagrams are thought to contribute the same magnitude to the
cross section as order $\alpha_s^2$ diagrams.  Measuring the cross section
as a function of $\Delta \phi$ provides information on the contributions of 
the leading order and next-to-leading order terms.

The event selection requires two central calorimeter towers
with significant $E_{\rm T}$ in association with two tracks in the central
tracking chamber with significant $p_{\rm T}$.  The calorimeter towers are
clustered, and the tracks are reconstructed with the Silicon Vertex
Trigger~\cite{svt}.  Events having two clusters with significant $E_{\rm T}$
associated with two tracks having significant displaced vertices are retained.
Jets are reconstructed with the Run I cone algorithm.  One or more
high-quality primary vertices are required.  Corrections are applied to the
jet energy scale as well as to the energies of $b$-jets (whose fragmentation,
for example,
is harder than that of light quarks).  The presence of two jets with 
significant $E_{\rm T}$, both $b$-tagged and with displaced secondary
vertices, is confirmed.  The data are compared to predictions by PYTHIA,
with Tune A; HERWIG; and MC@NLO~\cite{mcnlo} with the HERWIG parton shower
and the
underlying event generated by Jimmy 4.3~\cite{jimmy}.  The systematics are dominated
by the jet energy scale and range from 20 to 30\%.

\begin{figure}
\includegraphics[width=135mm]{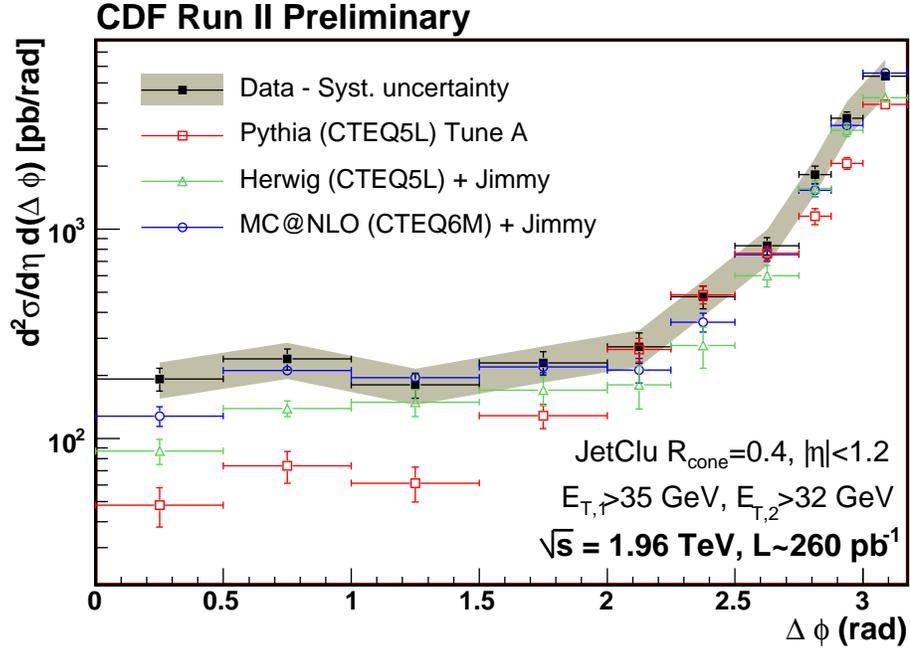}
\caption{The $b\overline b$ jet cross section as a function of the dijet
$\Delta\phi$ correlation.  Data are compared to hadron level cross sections
obtained using MC@NLO+JIMMY, PYTHIA, and HERWIG+JIMMY. The shaded area
represents the total systematic uncertainty on the data.\label{cross1}}
\end{figure}

For integrated luminosity 260 pb$^{-1}$, all three Monte Carlos show similar
agreement with data distibutions of $b\overline b$ production cross section
versus leading jet $E_{\rm T}$ and versus dijet invariant mass.  The cross
section as a function of $\Delta \phi$, however, is significantly better
modeled by MC@NLO, as shown in Figure~\ref{cross1}.  The peak at large
angles reflects flavor creation, and the excess at small angles is due
to higher order diagrams and multiple interactions.  The measured total
cross section for $|\eta_{12}|<1.2$, $E_{T1}>35$ GeV, and $E_{T2}>32$ GeV
is $\sigma=5664 \pm 168 \pm 1270$~pb.

\end{document}